\documentclass[pra,aps,twocolumn,showpacs,10pt]{revtex4-1}
\usepackage{amsmath,amssymb,graphicx,amsthm,dsfont,bm,color,ulem}

\begin{document}

\title{Effects of orbital angular momentum on few-cycle pulse shape: \\ Coupling between the temporal and angular momentum degrees of freedom}

\author{Miguel A. Porras}

\affiliation{Grupo de Sistemas Complejos, ETSIME, Universidad Polit\'ecnica de Madrid, Rios Rosas 21, 28003 Madrid, Spain}

\begin{abstract}
We describe the coupling between orbital angular momentum (OAM) and temporal degrees of freedom in pulsed Laguerre-Gauss beams. The effects of this coupling are to increase significantly the duration of few-cycle pulses when the OAM carried by the pulse is large, and to enhance spatiotemporal couplings. Contrary to other detrimental but retrievable effects such as spatial, group velocity and topological charge dispersions in these pulses, OAM-temporal coupling effects are unavoidable, and have therefore an impact in applications such as OAM-based optical communications, or in the generation of vortex-carrying attosecond pulses.
\end{abstract}

\maketitle

\noindent An ultrashort, few-cycle pulse shaped in space as a fundamental Gaussian beam does not substantially alters its temporal few-cycle shape when it is carefully focused  \cite{HOFF,PORRASF}. Coupling between the spatial and temporal degrees of freedom may be important in some circumstances \cite{AKTURK,BOR}, but their effects on pulse shape can be controlled and minimized so as not to affect significantly the temporal shape \cite{PORRASF,PORRASD}.

There is currently huge interest in few-cycle pulses carrying orbital angular momentum (OAM), usually shaped as Laguerre-Gauss (LG) beams. Reaching the few-cycle regime required to overcome technical difficulties such as the elimination of spatial, group velocity and topological charge dispersions introduced by fork gratings and spiral phase plates \cite{BEZUHANOV,ZEYLIKOVICH,TOKIZANE,SHVEDOV,RICHTER,YAMANE}, and now few-cycle pulsed LG-like beams are being employed in variety of fields such as strong-field physics to excite high harmonics and attosecond pulses carrying large OAM \cite{HERNANDEZ,GARIEPY,REGO}. Ultrashort pulses with high OAM are also fundamental objects in classical and quantum information science and technologies \cite{HUANG,MOLINA,FICKLER}. It is then fundamental to acquire an accurate knowledge of their propagation properties. Interestingly, the number of experimental works \cite{BEZUHANOV,ZEYLIKOVICH,TOKIZANE,SHVEDOV,RICHTER,YAMANE} far exceeds theoretical studies \cite{FENG,PARIENTE,LECKNER,CONTI,PORRAS1}.

For few-cycle vortex pulses there exist, in addition to the spatiotemporal coupling, a coupling between the OAM and the temporal degrees of freedom that makes the temporal shape strongly dependent on the OAM, as recently described for diffraction-free X-waves with OAM \cite{CONTI}. While the implementation these X-waves will benefit, e. g., optical communication technologies, current experiments employ or generate pulsed LG-like beams. For pulsed LG beams, it has recently been shown that there is a fundamental lower bound to the pulse duration for them to carry $l$ units of OAM \cite{PORRAS1}. Accordingly, with a certain available spectral bandwidth the synthesized pulse shape cannot be independent of the imprinted topological charge $l$. In this Letter we describe the effects of OAM-temporal coupling in ultrashort pulse shape. Pulses broaden as the OAM that is imposed to carry is higher, and previously known spatiotemporal coupling effects as red- and blue frequency shifts are enhanced via the OAM coupling. Thus, a few-cycle pulse in which a vortex is nested and is, in most cases, focused, does not retain its shape, but acquires a form that depends on OAM.

We consider the ultrashort pulsed beam formed by the superposition
\begin{equation}\label{AN}
E(r,\phi,z,t')= \frac{1}{\pi}\int_0^{\infty} \hat E_\omega(r,\phi,z)e^{-i\omega t'} d\omega\, ,
\end{equation}
of Laguerre-Gauss (LG) beams
\begin{equation}\label{LG}
\hat E_\omega(r,\phi,z) = \hat a_\omega \frac{e^{-i(|l|+1)\psi(z)}e^{-il\phi}}{\sqrt{1+\left(\frac{z}{z_R}\right)^2}}\left(\frac{\sqrt{2}r}{s_\omega(z)}\right)^{|l|}e^{\frac{i\omega r^2}{2cq(z)}} \, ,
\end{equation}
of same charge $l$, zero radial order, but different frequencies $\omega$ and weights $\hat a_\omega$. In the above relations $(r,\phi,z)$ are cylindrical coordinates, $t'=t-z/c$ is the local time, $c$ is the speed of light in vacuum, $q(z)=z-iz_R$ is the complex beam parameter, $\psi(z)=\tan^{-1}(z/z_R)$ Gouy's phase, and $s_\omega(z)=s_\omega\sqrt{1+(z/z_R)^2}$ and $s_\omega =\sqrt{2z_R c/\omega}$ are the Gaussian and Gaussian waist width. Among the many possible models of pulsed LG beams, we assume a frequency-independent Rayleigh distance $z_R$, or isodiffracting model \cite{PORRAS2,PORRAS3,FENG}. As shown recently, only in this case a few-cycle pulse retains its few-cycle shape during propagation regardless of the value of the topological charge, which is, on the other hand, the most desirable experimental situation. In this way, we also can discern the effects of OAM on pulse shape from those arising from propagation. In the isodiffracting model, the Gaussian width is inversely proportional to the square root of frequency, as in transverse modes of a stable two-mirror resonator. Additionally, the complex beam parameter is usually written as
\begin{equation}\label{Q}
\frac{1}{q(z)} = \frac{1}{R(z)} + i \frac{2c}{\omega s_\omega^2 (z)}\, ,
\end{equation}
where $1/R(z)= z+ z_R^2/z$ is the radius of curvature of the wave fronts.

\begin{figure}
\centering
\includegraphics[width=4.4cm]{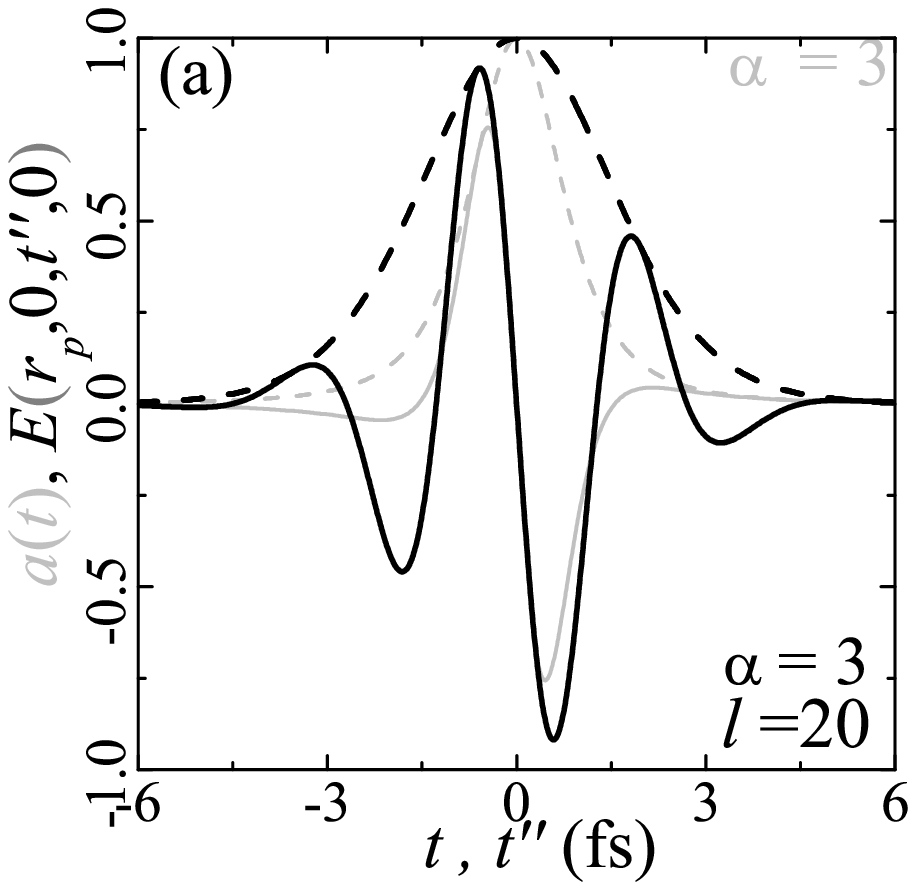}\includegraphics[width=4.4cm]{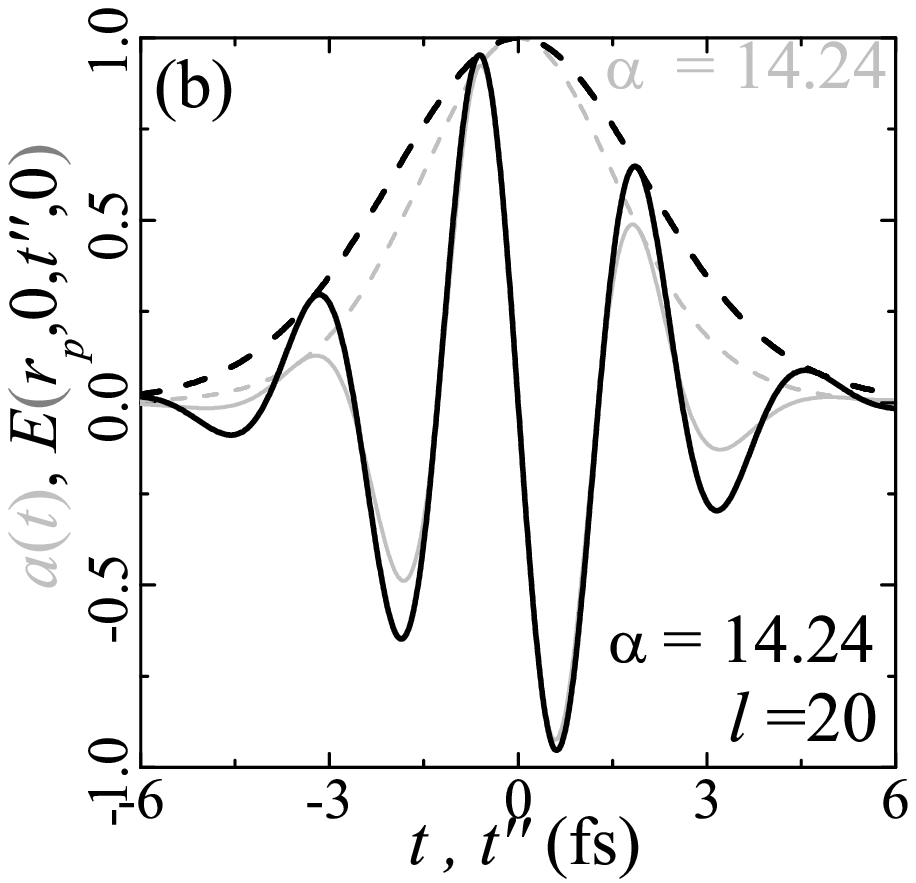}
\includegraphics[width=4.4cm]{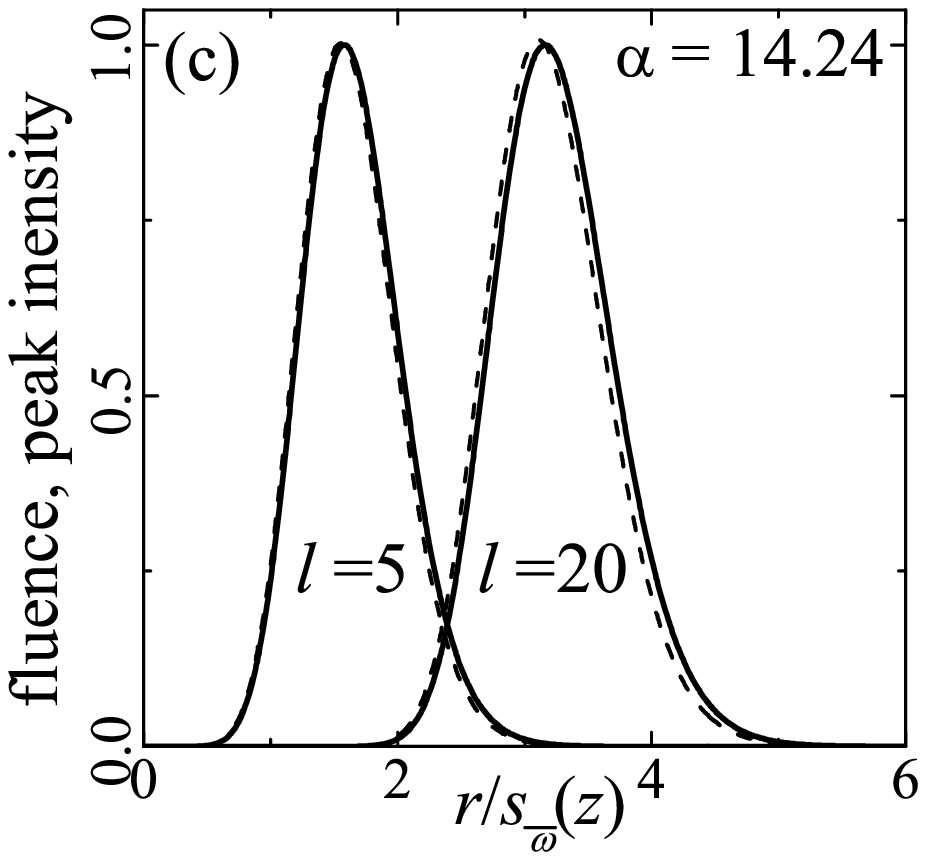}\includegraphics[width=4.4cm]{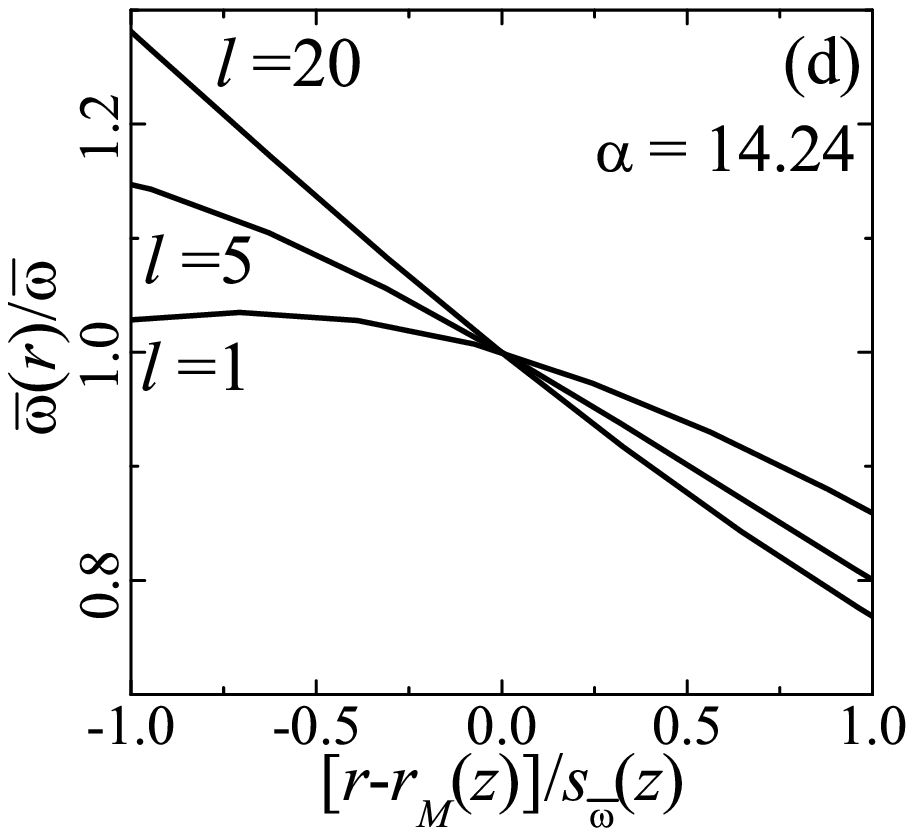}
\caption{\label{Fig1} (a) and (b) Gray curves: Pulse shape $\mbox{Re}\{a(t)\}$ (solid) and its modulus $|a(t)|$ (dashed curves) in (\ref{a}) with $\bar\omega=2.417$ fs$^{-1}$ ($780$ nm wavelength) with the spectrum (\ref{PE}) with the indicated values of $\alpha$ and carrier-envelope phase $\Phi=\pi/2$. Black curves: broadened pulse shape at the bright ring $r_M=s_{\bar\omega}\sqrt{|l|/2}$ at $z=0$ of the pulsed LG beam of charge $l=20$ synthesized with the PE spectrum (\ref{PE}) with the indicated values of $\alpha$ and $\Phi=\pi/2$. (c) Radial profile of fluence and peak intensity (at $t^{\prime\prime}=0$), normalized to their values at $r_M(z)=s_{\bar\omega}(z)\sqrt{|l|/2}$ for a single-cycle pulse and various $l$. (d)Mean frequency $\bar\omega(r)$ (relative to $\bar\omega$) within the thickness of the bright ring with center at $r_M(z)=s_{\bar\omega}(z)\sqrt{|l|/2}$ for a single-cycle pulse and various $l$. }
\end{figure}

As a flexible model we consider the weights
\begin{equation}\label{PE}
\hat a_\omega= \frac{\pi}{\Gamma(\alpha+1/2)}\frac{\alpha^{\alpha+1/2}}{\bar\omega}\left(\frac{\omega}{\bar\omega}\right)^{\alpha-1/2}
e^{-\frac{\omega}{\bar\omega}\alpha}e^{-i\Phi}\,,
\end{equation}
where $\alpha > 1/2$ and $\Phi$ is an arbitrary phase. Equation (\ref{PE}) is an appropriate way to write the so-called power-exponential (PE) spectrum \cite{CONTI,PORRAS3,FENG} in which the mean frequency $\bar\omega=\int_0^\infty |\hat a_\omega|^2\omega d\omega/ \int_0^\infty |\hat a_\omega|^2d\omega$ appears explicitly, and the Gaussian-equivalent half bandwidth $\Delta \omega ^2 = 4 \int_0^\infty |\hat a_\omega|^2(\omega-\bar\omega)^2 d\omega/ \int_0^\infty |\hat a_\omega|^2d\omega$ (yielding approximately the $1/e^2$ decay of $|\hat a_\omega|^2$) is given by $\Delta\omega=\bar\omega\sqrt{2/\alpha}$. The pulse shape with the PE spectrum
\begin{equation}\label{a}
a(t)=\frac{1}{\pi}\int_0^\infty \hat a_\omega e^{-i\omega t}d\omega = \left(\frac{-i\alpha}{\bar\omega t -i\alpha}\right)^{\alpha +1/2}e^{-i\Phi}
\end{equation}
is characterized by increasing number of oscillations with increasing $\alpha$, approaching the Gaussian-enveloped pulse $a(t)\simeq e^{-t^2/\Delta t^2}\cos(\bar\omega t +\Phi)$ of Gaussian half duration $\Delta t= 2/\Delta\omega = \sqrt{2\alpha}/\bar\omega$ and carrier envelope phase $\Phi$. To fix ideas the full width at half maximum (FWHM) of $|a(t)|^2$ is $2\sqrt{\alpha\ln 2}/\bar\omega$, which equated to $N(2\pi/\bar\omega)$ yields the number of oscillations $N= \sqrt{\alpha\ln 2}/\pi$ within the FWHM. For example, a single-cycle ($N=1$) Gaussian-like pulse corresponds to $\alpha= 14.24$. The gray curves in Figs. \ref{Fig1}(a) and \ref{Fig1}(b) are examples with $\alpha=3$ and the single-cycle pulse with $\alpha=14.24$. For $\alpha\ge 14.24$ the pulse $a(t)$ has a physically meaningful carrier frequency $\omega_0=\bar\omega$ and complex envelope $A(t)=a(t)e^{i\omega_0 t}$ \cite{BRABEC}, but we also consider sub-cycle pulses.

Integral (\ref{AN}) with (\ref{LG}) and (\ref{PE}) can be carrier out, yielding the closed-form expression of the pulsed LG beam
\begin{eqnarray}\label{ADAPT}
E(r,\phi,z,t') &=& \frac{e^{-i(|l|+1)\psi(z)}e^{-il\phi}}{\sqrt{1+\left(\frac{z}{z_R}\right)^2}}\left(\sqrt{\frac{2}{|l|}}\frac{r}{s_{\bar\omega}(z)}\right)^{|l|} \nonumber \\
&\times & \left[\frac{-i\left(\alpha+ \frac{|l|}{2}\right)}{\bar\omega\left(t'-\frac{r^2}{2cq(z)}\right)-i\alpha}\right]^{\alpha+\frac{|l|}{2}+\frac{1}{2}} e^{-i\Phi}\,,
\end{eqnarray}
where $s_{\bar\omega}(z)$ is the Gaussian width for the mean frequency, and where the value of $|E|$ at the waist or focus $z=0$, at the ring $r_M=s_{\bar\omega}\sqrt{|l|/2}$ and at the time of pulse peak $t'=0$ is adjusted to unity. We first note that (\ref{ADAPT}) has no singularities, approaches zero as $r\rightarrow\infty$ as $1/r^{2\alpha+1}$, and carrying then finite instantaneous power and finite energy for $\alpha>1/2$. Using  (\ref{Q}) particularized at the frequency $\bar\omega$, the denominator in the temporal factor of (\ref{ADAPT}) can be written as $\omega_0t^{\prime\prime}-i\left[\alpha+r^2/s_{\bar\omega}^2(z)\right]$, where $t^{\prime\prime}\equiv t'- r^2/2cR(z)$. We then see that the pulse front $t^{\prime\prime}=0$ matches the common spherical wave fronts of radius $R(z)$ of all LG beams components, as for the fundamental ($l=0$) isodiffracting pulsed Gaussian beam \cite{PORRAS2}. A slow detector will measure the time-integrated intensity, fluence or energy density, given by $F(r,z) =\int_{-\infty}^\infty (\mbox{Re} E)^2 dt = (1/2)\int_{-\infty}^\infty |E|^2 dt = (1/\pi)\int_0^\infty |\hat E_\omega|^2 d\omega$, which can be evaluated to be
\begin{equation}\label{fluence}
F(r,z) = \frac{1}{1+\left(\frac{z}{z_R}\right)^2}\left[\frac{2}{|l|}\frac{r^2}{s^2_{\bar\omega}(z)}\right]^l\left[\frac{\alpha+\frac{|l|}{2}}{\alpha + \frac{r^2}{s_{\bar\omega}^2(z)}}\right]^{2\alpha+|l|}\,.
\end{equation}
The fluence features a maximum at a ring with center at $r_M(z)=s_{\bar\omega}(z)\sqrt{|l|/2}$ at each $z$, of approximate width $s_{\bar\omega}(z)$. In (\ref{fluence}), the value of the energy density is set to unity at this ring at the waist $z=0$. Figure \ref{Fig1}(c) shows fluence profiles for several values of $l$ peaking at $r_M(z)$, along with the respective profiles of pulse peak intensities, or squared modulus of (\ref{ADAPT}) at $t^{\prime\prime}=0$.

Being $r_M(z)=s_{\bar\omega}(z)\sqrt{|l|/2}$ the center of the bright ring, it is relevant to analyze the pulse shape at this radius. At $r_M(z)=s_{\bar\omega}(z)\sqrt{|l|/2}$, (\ref{ADAPT}) yields
\begin{equation}\label{PR}
E= D\left[\frac{-i\left(\alpha + \frac{|l|}{2}\right)}{\bar\omega t^{\prime\prime}- i\left(\alpha+ \frac{|l|}{2}\right)}\right]^{\alpha +\frac{|l|}{2}+ \frac{1}{2}}e^{-i\Phi},
\end{equation}
where $D(z,\phi)=e^{-i(|l|+1)\psi(z)}e^{-il\phi}/\sqrt{1+\left(z/z_R\right)^2}$.
The pulse shape is seen to be the same as $a(t)$ in (\ref{a}) but with $\alpha$ replaced with $\alpha + |l|/2$, and peaking at $t^{\prime\prime}=0$. Thus, with the same frequency spectrum (\ref{PE}), the synthesized pulse is different with different values of $l$, becoming longer as the carried OAM is higher. Specifically, the bandwidth is $\Delta\omega_l  =\bar\omega\sqrt{2/(\alpha+ |l|/2)}$; hence the duration is $\Delta t_l  = 2/\Delta\omega_l  = \sqrt{2(\alpha + |l|/2)}/\bar\omega$, and the number of oscillations is $N_l= \sqrt{(\alpha+ |l|/2)\ln 2}/\pi$. In other words, the pulse adapts itself to carry the OAM by increasing its duration and number of oscillations by a factor $\sqrt{1+ |l|/2\alpha}$ compared to the minimum number of oscillations $N=\sqrt{\alpha\ln 2}/\pi$ achievable with the available bandwidth $\Delta\omega$ in a fundamental pulsed Gaussian beam. In terms of $N$ (to remove the less physical parameter $\alpha$), the number of oscillations in the pulsed LG beam is
\begin{equation}\label{NUMBER}
N_l=\sqrt{N^2 + \frac{\ln 2}{2\pi^2}|l|}\,.
\end{equation}
The black curves in Figs. \ref{Fig1}(a) and \ref{Fig1}(b) represent two examples of elongated pulse shapes at the bright ring in order to transport $l=20$ units of OAM. Coupling between the temporal and OAM degrees of freedom is more pronounced as the pulse is shorter and OAM is higher. On the opposite side, coupling vanishes in the monochromatic limit. Pulse reshaping is a natural consequence of the fundamental restriction between the OAM and the pulse bandwidth or duration, $4\bar\omega^2/\Delta\omega_l^2>|l|$ or $\Delta t_l \bar\omega >\sqrt{|l|}$, at the bright ring of the pulsed LG beam, recently found in Ref. \cite{PORRAS1}. For (\ref{PR}) indeed $4\bar\omega^2/\Delta\omega_l^2= |l|+2\alpha>|l|$ and $\Delta t_l \bar\omega=\sqrt{|l|+2\alpha}>\sqrt{|l|}$ are satisfied with any $\alpha$ because of the dependence of the pulse shape on $l$.

At other radii, simple manipulation of (\ref{ADAPT}) yields
\begin{eqnarray}
E &=& D \left(\sqrt{\frac{2}{|l|}}\frac{r}{s_{\bar\omega}(z)}\right)^{|l|}\left(\frac{\alpha + \frac{|l|}{2}}{\alpha + \frac{r^2}{s^2_{\bar\omega}(z)}}\right)^{\alpha+ \frac{|l|}{2}+\frac{1}{2}} \nonumber \\
&\times&\left[\frac{-i\left(\alpha + \frac{|l|}{2}\right)}{\bar\omega(r) t^{\prime\prime}- i\left(\alpha+ \frac{|l|}{2}\right)}\right]^{\alpha +\frac{|l|}{2}+ \frac{1}{2}}e^{-i\Phi}\, ,
\end{eqnarray}
where
\begin{equation}
\bar\omega(r)= \frac{\alpha + \frac{|l|}{2}}{\alpha + \frac{r^2}{s_{\bar\omega}^2(z)}}\bar\omega
\end{equation}
The pulse shape is then the same as at the bright ring, but scaled in time to oscillate at the mean frequency $\bar\omega(r)$. In particular the number of oscillations $N_l= \sqrt{(\alpha+ |l|/2)\ln 2}/\pi$ is the same as at the bright ring. The frequency is red-shifted (blue-shifted) outside (inside) the bright ring. Similar effect of spatiotemporal coupling has previously been described for the fundamental Gaussian beam \cite{PORRAS2,PORRAS3}, but with $l\neq 0$ they are much more pronounced; for instance, while red-shift within the beam width $s_{\bar\omega}(z)$ of a single-cycle fundamental Gaussian beam ($l=0$) is almost negligible, with increasing $|l|$ the blue- and red-shifts within the ring thickness can reach 20\% for moderate magnitudes of $l$, as seen in Fig. \ref{Fig1}(d).

\begin{figure}[t]
\centering
\includegraphics*[height=4.2cm]{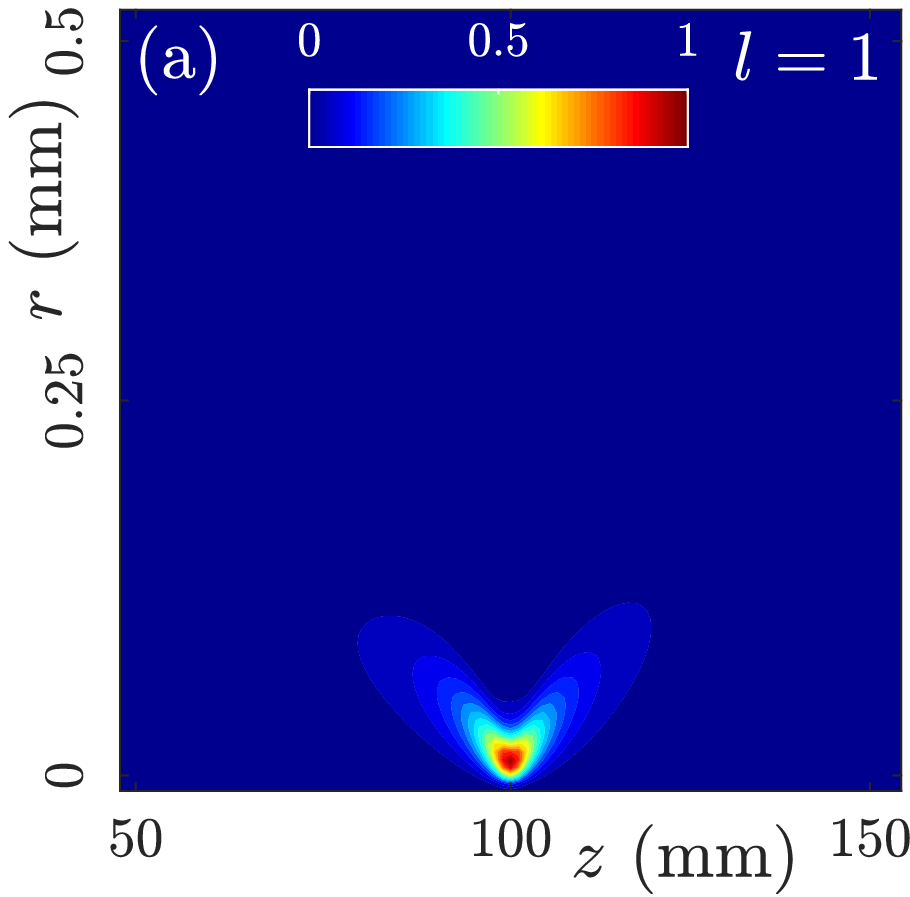}\includegraphics*[height=4.2cm]{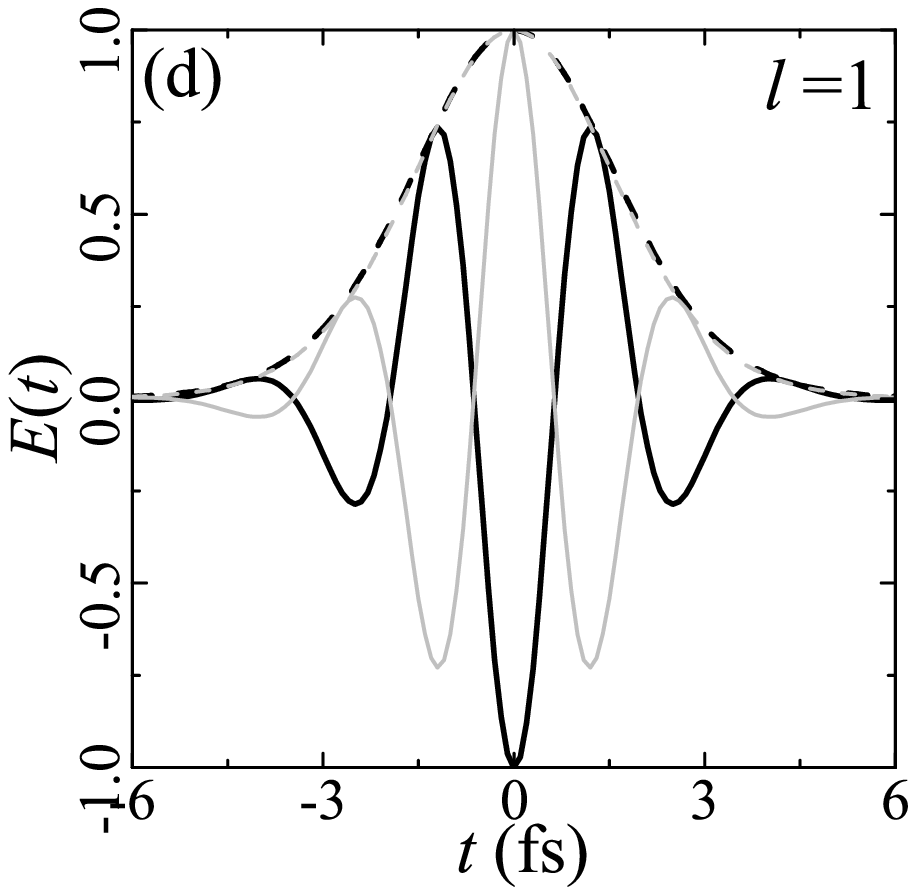}
\includegraphics*[height=4.2cm]{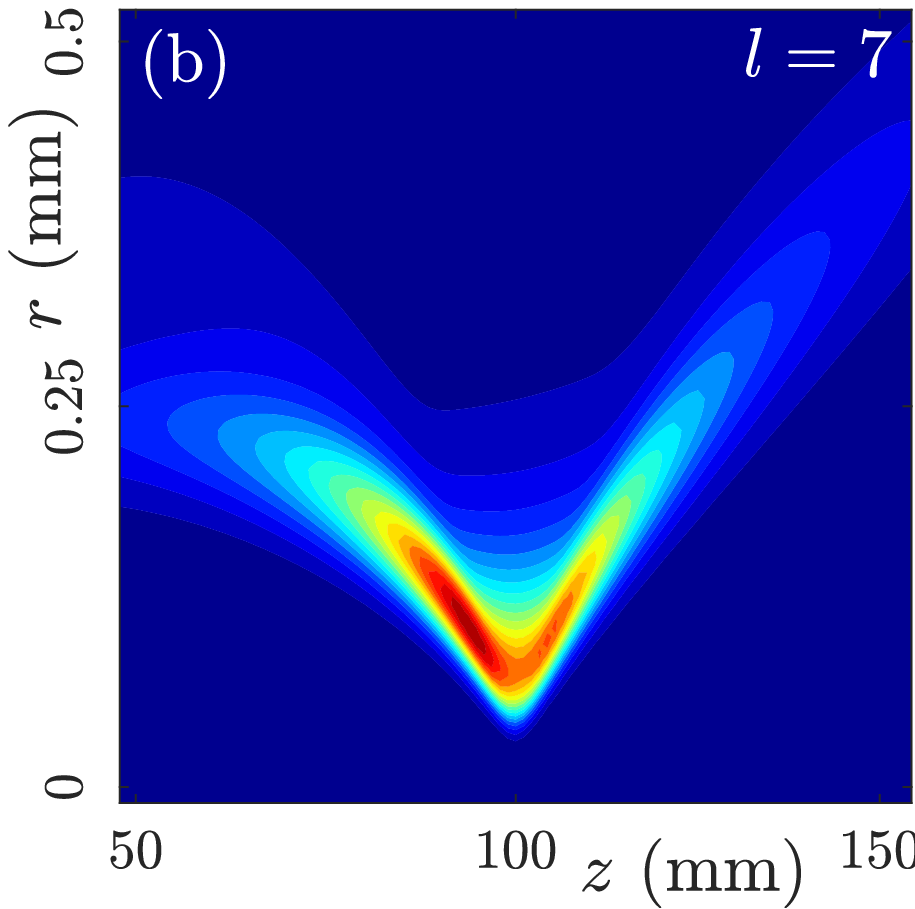}\includegraphics*[height=4.2cm]{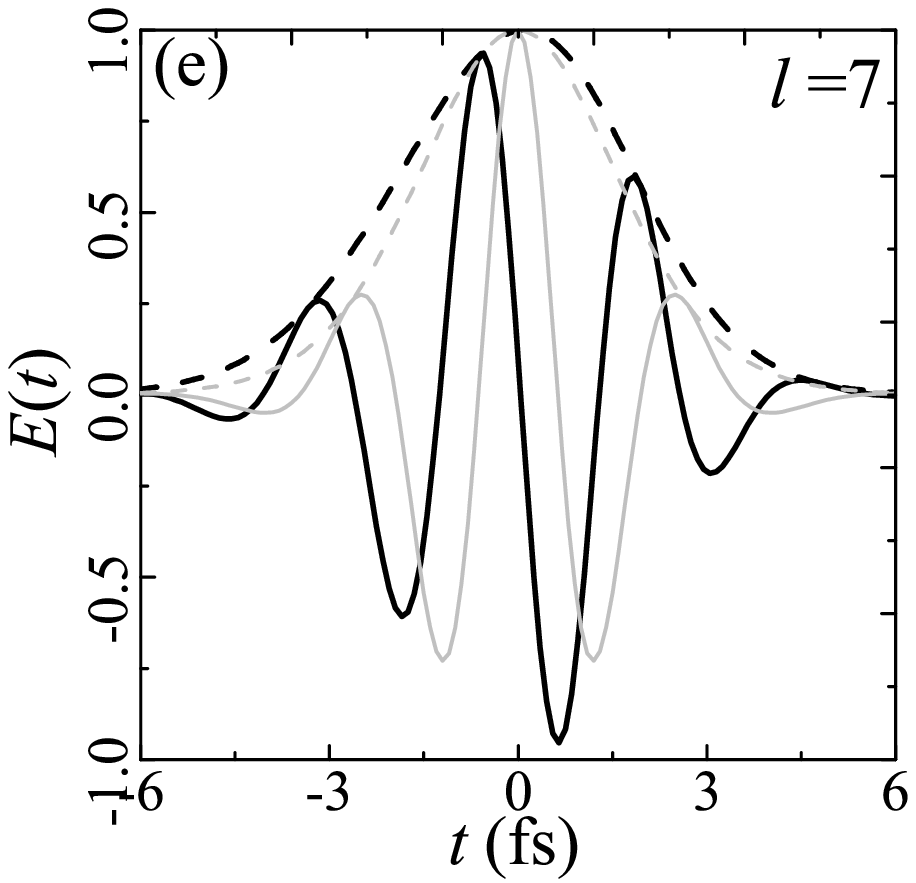}
\includegraphics*[height=4.2cm]{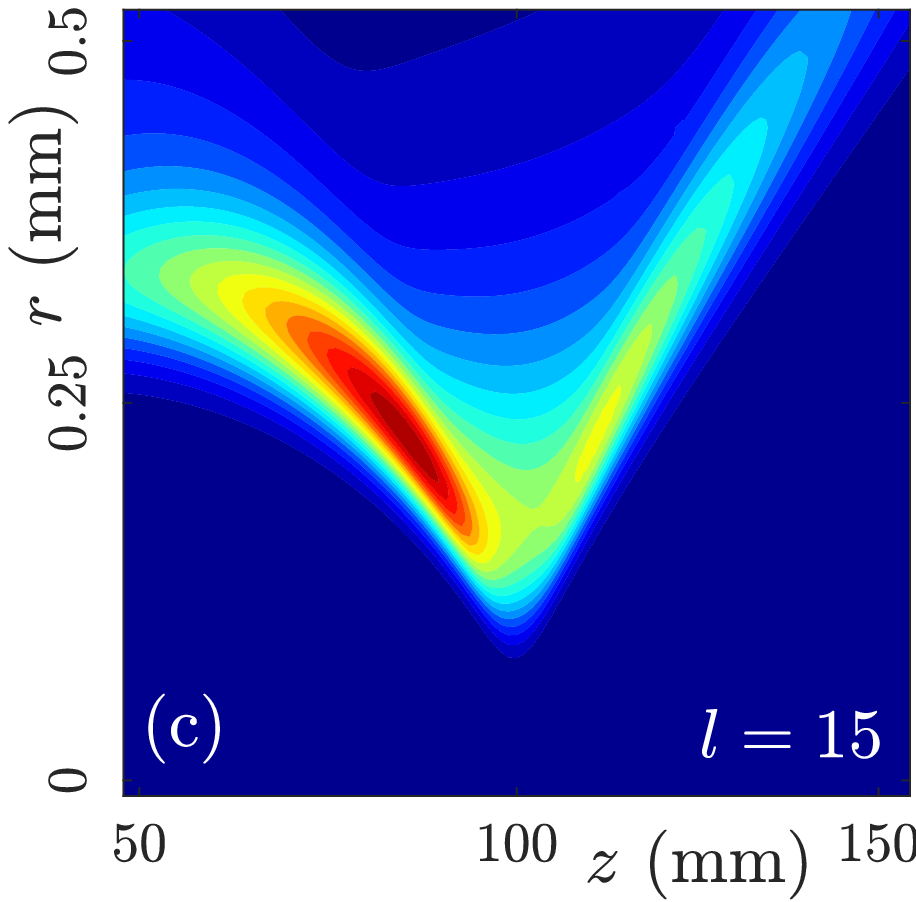}\includegraphics*[height=4.2cm]{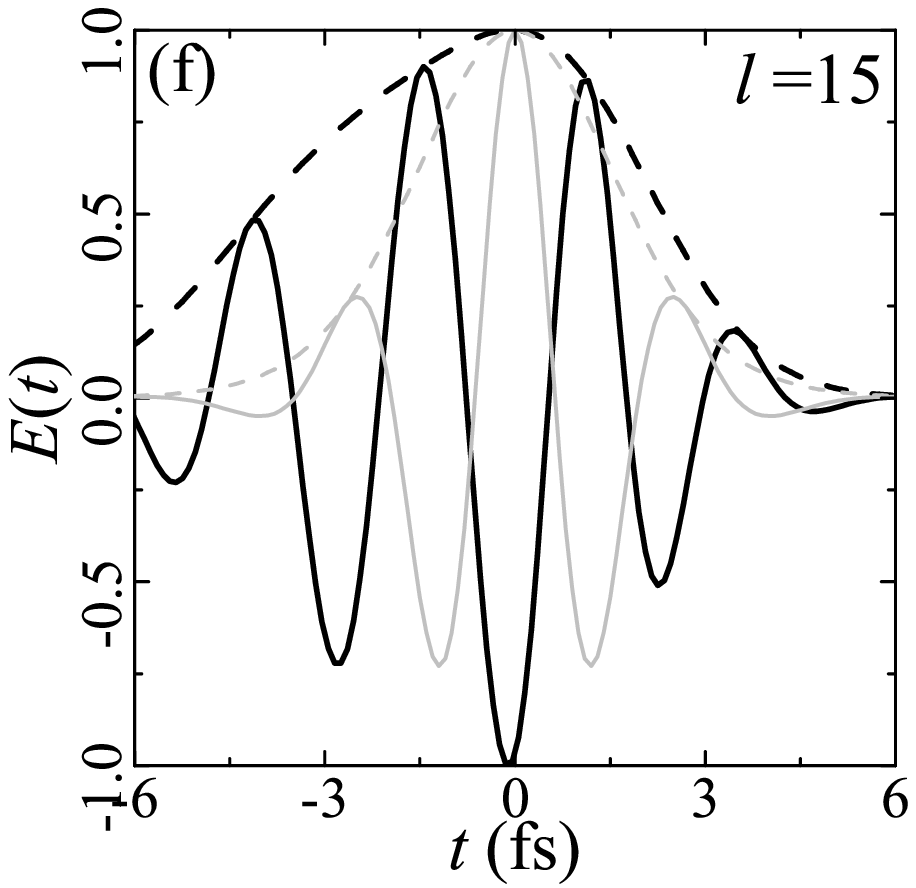}

\caption{\label{Fig2} Focusing of a single-cycle ($\alpha=14.24$) collimated Gaussian beam with nested punctual vortex of the indicated topological charges. The Rayleigh distance of the input Gaussian beam is $z_{R,L}=5000$ mm, the carrier frequency is $\bar\omega=2.417$ fs$^{-1}$ ($780$ nm wavelength), yielding Gaussian width at the carrier frequency $s_{\bar\omega,L}=1.114$ mm, and the focal length is $f=100$ mm. (a-c) Fluence distributions about the focal region. (d-f) Gray curves: Input pulse shape (solid) and its envelope (dashed). Black curves: Focused pulse shape (solid) and its envelope (dashed) at the respective points of maximum fluence. The focused pulses oscillate at the same carrier frequency $\bar\omega$ as the input pulse but are longer as the magnitude of the charge $l$ is larger.}
\end{figure}

The above couplings between the OAM and temporal degrees of freedom are expected to be observable in experiments involving focusing of few-cycle pulses, particularly if they are forced to carry high OAM. For example, setting $z_R=f^2/z_{R,L}$ and $\hat a_\omega=(-iz_{R,L}/f)\hat a_{\omega,L}$, (\ref{ADAPT}) represents the focused few-cycle pulse about the focus (at $z=0$) when an ideal focusing system of focal length $f$ is illuminated by collimated pulsed LG beam of monochromatic LG components $\hat a_{\omega,L}(\sqrt{2}r/s_{\omega,L})^{|l|}e^{-r^2/s_{\omega,L}^2}e^{-il\phi}$ of Rayleigh range $z_{R,L}$, Gaussian spot size $s_{\omega,L}=\sqrt{2z_{R,L} c/\omega}$ and spectral amplitude $\hat a_{\omega,L}$ given by (\ref{PE}).

In a less idealized situation, we consider the usual arrangement in which a collimated, few-cycle Gaussian beam ($l=0$) illuminates a system that imprints the azimuthal phase $e^{il\phi}$, as a spiral phase plate, a fork grating or a spatial light modulator. The vortex pulsed beam is then focused by a focusing system of focal length $f$ such as a spherical or parabolic mirror. The monochromatic components immediately after the mirror are then $\hat E_\omega(r,\phi)= \hat a_{\omega,L} e^{-r^2/s_{\omega,L}^2}e^{-il\phi}e^{-i\omega r^2/2cf}$, where $s_{\omega,L}=\sqrt{2z_{R,L} c/\omega}$ is the Gaussian width on the focusing system, $z_{R,L}$ its Rayleigh distance, and $\hat a_{\omega,L}$ is given by (\ref{PE}). The temporal shape of the pulse in front of the focusing system is then independent of $l$ and given at $r=0$ by (\ref{a}), e. g., the single-cycle pulse with $\alpha=14.24$. The propagated monochromatic components of a punctual vortex placed in a Gaussian beam have been evaluated from Fresnel diffraction integral \cite{SACKS,BEKSHAEV,ANZOLIN}, are known to differ significantly from LG beams, and are given by
\begin{eqnarray}
\hat E_\omega(r,\phi,z) &=& i^{l-1} \sqrt{\frac{\pi}{2}}\hat a_\omega \frac{p}{Z} e^{il\phi}e^{\frac{i\omega r^2}{2cZ}} \nonumber \\
&\times & \gamma^{\frac{1}{2}}e^{-\gamma} \left[I_{\frac{l-1}{2}}(\gamma) - I_{\frac{l+1}{2}}(\gamma)\right] \, ,
\end{eqnarray}
where $\gamma=\omega r^2 p/4cZ^2$, $p=[1/z_{R,L} -i(1/f-1/Z)]^{-1}$, $Z=z+f$ is an axial coordinate with origin on the focusing system ($z=0$ at the geometrical focus), and $I_m(\cdot)$ is the modified Bessel function of the first kind. The pulsed beam in time domain is then obtained as $E(r,\phi,z,t')=(1/\pi)\int_0^{\infty}E_\omega(r,\phi,z)e^{-i\omega t'}d\omega$. Figures \ref{Fig2}(a-c) show fluence distributions in the focal region for a single-cycle pulse in front of the focusing system and for increasing values of the imprinted topological charge $l$, and Figs. \ref{Fig2}(d-f) show corresponding pulse shapes and envelopes at the points $(r,z)$ where the fluence takes its absolute maximum vale, i. e., at the effective best focus. With $l=1$, the focused pulse continues to be substantially the same as the single-cycle pulse that illuminated the system, but with increasing $l$ the pulse features increasing duration and number of oscillations without changing its carrier frequency, as in the pulsed LG model. This effect may be more or less pronounced with different beam geometries (e. g., with flat-top input beam), but being a consequence of the restriction $\Delta t_l \bar\omega >\sqrt{|l|}$ \cite{PORRAS1}, it is unavoidable even if other deteriorating effects such as group velocity, angular and topological charge dispersions introduced by fork gratings or spiral phase plates are compensated \cite{BEZUHANOV,ZEYLIKOVICH,TOKIZANE,SHVEDOV,RICHTER,YAMANE}, as we have assumed in the above example.

In conclusion, we have demonstrated the existence of a coupling between the temporal and OAM degrees of freedom in few-cycle or sub-cycle pulses in the usual beam geometry in experiments involving diffraction and focusing. Coupling effects differ substantially from those described in \cite{CONTI} diffraction-free X-waves. With the same weights of the monochromatic spectral components the pulse shape adapts to carry the imposed topological charge, the coupling effects being more pronounced as the pulse is shorter and the OAM is higher. Few-cycle pulses with high OAM are currently being produced in experiments \cite{HERNANDEZ,GARIEPY,REGO}. The coupling effects on pulse shape must therefore be taken into account for a correct design and interpretation of the experiments.

The author acknowledges support from Projects of the Spanish Ministerio de Econom\'{\i}a y Competitividad No. MTM2015-63914-P, and No. FIS2017-87360-P.

\end{document}